\pdfoutput=1
\documentclass[12pt]{article}
\usepackage{times}
\usepackage{graphics}
\usepackage{graphicx}
\usepackage{cite}
\newcommand{\Sc}{Schr\"odinger equation}
\newcommand{\n}{\mbox{\boldmath $n$}}
\newcommand{\si}{\mbox{\boldmath $\sigma$}}
\newcommand{\rr}{\mbox{\boldmath $r$}}
\newcommand{\B}{\mbox{\boldmath $B$}}
\newcommand{\mb}{\mbox{\boldmath $\mu$}}
\newcommand{\eref}[1]{(\ref{#1})}
\newcommand{\hr}{\mbox{\boldmath$\hat{r}$}}
\newcommand{\hw}{\mbox{\boldmath$\hat{w}$}}
\newcommand{\htt}{\mbox{\boldmath$\hat{t}$}}
\newcommand{\hk}{\mbox{\boldmath $\hat{k}$}}
\newcommand{\hPs}{\mbox{\boldmath $\hat{\Psi}$}}
\newcommand{\hR}{\mbox{\boldmath $\hat{R}$}}
\newcommand{\shk}{\mbox{\small\boldmath $\hat{k}$}}
\newcommand\fr{\displaystyle\frac}
\newcommand\lt{\left}
\newcommand\rt{\right}
\newcommand{\hG}{\mbox{\boldmath $\hat{G}$}}
\newcommand{\hI}{\mbox{\boldmath $\hat{I}$}}
\begin{document}
\large
\date{ }

\begin{center}
{\Large Limits on a nucleon-nucleon monopole-dipole coupling from
spin relaxation of polarized ultra-cold neutrons in traps}

\vskip 0.7cm

V. K. Ignatovich and Yu. N. Pokotilovski \footnote{e-mail:
pokot@nf.jinr.ru}

\vskip 0.7cm
            Joint Institute for Nuclear Research\\
              141980 Dubna, Moscow region, Russia\\
\vskip 0.7cm

{\bf Abstract\\}

\begin{minipage}{130mm}

\vskip 0.7cm
 A new limit is presented on the axion-like monopole-dipole
P,T-non-invariant interaction in a range $(10^{-4} - 1)$\,cm.
 The spin-dependent nucleon-nucleon potential between neutrons
and nucleons of the walls of the cavity containing ultra-cold
neutrons should affect the neutron depolarization probability at
their reflection from the walls.
 The limit is obtained from existing data on the ultra-cold neutron
depolarization probability per one collision with the walls.

\end{minipage}
\end{center}

\vskip 0.3cm


\vskip 0.2cm


\vskip 0.6cm

\section{Introduction}
 Hypothetical pseudoscalar particle -- axion offers a window to probe very
small coupling and very high energy scales \cite{ax}.

 Axion, according to later modifications of the primary model \cite{ax},
has mass in a very large range:  $(10^{-12}<m_{a}<10^{6})$\,eV.
 Current algebra technique is used to relate the masses and coupling constants
of the axion and neutral pion: $m_{a}=(f_{\pi}m_{\pi}/f_{a})\sqrt{z}/(z+1)$,
where $z=m_{u}/m_{d}=0.56$, $f_{\pi}\approx 93$\,MeV, $m_{\pi}=135$\,MeV,
so that $m_{a}\approx (0.6\times 10^{10}$\,GeV/$f_{a})$\,meV.
 Here $f_{a}$ is the scale of Peccei-Quinn symmetry breaking.
 The axion coupling to fermions can be in general represented as
$g_{aff}=C_{f}m_{f}/f_{a}$, where $C_{f}$ is the model dependent factor
\cite{PDG,rev}.

 Early reactor, beam-dump, weak decays, and nuclear transition experiments
have placed lower limits on the axion mass.
 Stringent limits, especially from the lower side of the axion mass range,
have been set on the existence of axion using astrophysical and cosmological
arguments \cite{win,PDG}.
 These more recent constraints limit the axion mass to
$(10^{-5}<m_{a}<10^{-3})$\,eV with correspondingly very
small coupling constants to quarks and photon \cite{PDG,rev,win}.
 Although these limits are more stringent than can be reached in laboratory
experiments, it is of interest to try to constrain the axion as much as
possible using laboratory means.
 Interpretation of laboratory experiments depend on less number of assumptions
than the constraints inferred from astrophysical and cosmological observations
and calculations.
 The laboratory experiments performed or proposed so far are rather diverse
and employ a variety of detection techniques.
 Axion is still one of the candidates for the cold dark matter of the Universe
\cite{dark}.
 Some recent reviews are \cite{PDG,rev}.

\section{Monopole-dipole interaction potential}
 Axions mediate a P- and T-odd  monopole-dipole interaction potential between
spin and matter (polarized and unpolarized nucleons) \cite{Mood}:
\begin{equation}\label{1}
U({\bf r})=(\si\cdot\n)\frac{g_{s}g_{p}\hbar^2}{8\pi m_{n}}
\Bigl(\frac{1}{\Lambda r}+\frac{1}{r^{2}}\Bigr)e^{-r/\Lambda}=
(\si\cdot\n)V_{ps}(r),
\end{equation}
where $g_{s}$ and $g_{p}$ are dimensionless coupling constants of the scalar
and pseudoscalar vertices (unpolarized and polarized particles), $m_{n}$ is
the nucleon mass at the polarized vertex, $\si$ is vector of the Pauli
matrices related to the nucleon spin, $r$ is the distance between the
nucleons, $\Lambda=\hbar/m_{a}c$ is the range of the force,
$m_{a}$ is the axion mass, and $ \n=\rr/r$ is the unit vector.

 The microscopic potential \eref{1} between two nucleons creates a
macroscopic potential between nucleon and matter.
 If the matter is represented by a layer of thickness $d$, the neutron
interaction with it is described by the potential
\begin{equation}\label{2}
V(x)=(\si\cdot\n)g_{s}g_{p}\frac{\hbar^{2}N\Lambda}{4m_{n}}e^{-x/\Lambda}
(1-e^{-d/\Lambda})=V_{0}(\si\cdot\n)e^{-x/\Lambda}(1-e^{-d/\Lambda}),
\end{equation}
where $x$ is the distance to the interface along the normal
(unit vector $\n$), $N$ is the nucleon density in the layer, and in the
last equality we defined
\begin{equation}\label{3}
V_{0}=g_{s}g_{p}\frac{\hbar^{2}N\Lambda}{4m_{n}}.
\end{equation}
 We will deal here only with ultra-cold neutrons (UCN) or total reflection
of higher energy neutrons, and because of that we do not consider how the
potential changes, when a neutron enters the matter.

 Several laboratory searches provided constraints on axion-like coupling
in the macroscopic range $\Lambda >0.1$\,cm \cite{PDG,Hamm}.
 The limit in the $\Lambda$--range $(10^{-4}-1)$\,cm was established in the
Stern-Gerlach type experiment in which UCN were transmitted through a slit
between a horizontal mirror and absorber \cite{grav}.
 The obtained limit for the value $g_{s}g_{p}$ was $\sim 10^{-15}$ at
$\Lambda=10^{-2}$\,cm, that corresponds to the value of the monopole-dipole
interaction potential at the surface of the mirror $V_0\sim 10^{-3}$\,neV.
 This value is equivalent to an effective magnetic interaction $-\mb\B$ of the
neutron with magnetic field $B\sim 0.2$\,G.
 It is estimated that in future ultra-cold neutron Stern-Gerlach type
experiments sensitivity will be improved by orders of magnitude~\cite{grav1}.
 A better sensitivity is also expected in a proposed experiment on the
ultra-cold neutron magnetic resonance frequency shift~\cite{Zimm}.

 We consider here what limits on $g_{s}g_{p}$ in the range
$\Lambda\div(10^{-4}-1)$\,cm can be extracted from depolarization
of UCN in storage traps.
 Depolarization can be expected because the particle spin interaction with
axion field $U_{ax}=(\si\cdot\n)V$ is similar to magnetic interaction
$U_{magn}=|\mu|(\si\cdot\B)$ and corresponding pseudo-magnetic field in
general is not collinear to the neutron polarization.

 Depolarization was already considered earlier in the paper~\cite{Ser},
however it was estimated there semiclassically as the neutron spin rotation in
the interaction region $\Lambda$ in vicinity of the reflecting wall, which is
not sufficiently rigorous.
 Here we calculate the depolarization probability at a single collision with
the wall with distorted wave Born approximation (DWBA) perturbation theory.

\section{Constraints from the ultra-cold neutron depolarization}

 In fact, depolarization of UCN in traps can be attributed to many factors.
 In between them are inhomogeneities of the internal magnetic field and
presence of magnetic impurities on the walls.
 We use here the most conservative estimate attributing all the
neutron depolarization to the hypothetical axion-like interaction.

 Let's consider a nonmagnetic semi-infinitely thick wall with optical
potential $u$ at $x>0$, external homogeneous field $\B$ parallel to $z$-axis,
and the axion pseudo magnetic field $b(x)=b_0\exp(x/\Lambda)$ parallel to
$x$-axis.
 We want to calculate spin-flip reflectivity of a neutron initial polarization
along the $\B$ field, taking the axion field as a perturbation.
 For that we need to solve one dimensional stationary \Sc, which can be
represented in the form
\begin{equation}\label{a}
[d^2/dx^2+k^2-B\sigma_z-b(x)\sigma_x\Theta(x<0)-u\Theta(x>0)]|\Psi(x)\rangle=0,
\end{equation}
where $\sigma_{x,z}$ are the well known Pauli matrices, $\Theta(x)$ is a step
function equal to unity, when inequality in its argument is satisfied, and to
zero otherwise, $k$ is normal component in vacuum of the wave vector of the
incident particle, and the magnetic fields include factor $2m|\mu|/\hbar^2$,
which contains neutron mass $m$ and magnetic moment $\mu$.
 To find a spinor solution $|\Psi(x)\rangle$ of \eref{a} we need to define an
incident wave $|\psi_0(x)\rangle$.
 In general we can define it as
\begin{equation}\label{a1}
|\psi_0(x)\rangle=\exp(i\hk x)|\xi\rangle,
\end{equation}
where $\hk=\sqrt{k^2-B\sigma_z}$, and $|\xi\rangle$ is an arbitrary spin
state, which is a superposition
$|\xi\rangle=\alpha|\xi_u\rangle+\beta|\xi_d\rangle$ of states
$|\xi_{u,d}\rangle$ that are eigen spinors of the matrix
$\sigma_z$: $\sigma_z|\xi_{u,d}\rangle=\pm|\xi_{u,d}\rangle$.

 A non-perturbed solution of \eref{a} is
\begin{equation}\label{a2}
|\Psi_0(x)\rangle=\lt\{\Theta(x<0)[e^{i\shk x}+e^{-i\shk
x}\hr(k)]+\Theta(x>0)\htt(k)e^{i\shk' x}\rt\}|\xi\rangle,
\end{equation}
where $\hk'=\sqrt{k^2-B\sigma_z-u}$, reflection, $\hr$, and transmission,
$\htt$, matrices determined (see, for instance~\cite{ir}) from matching
conditions at the interface are
\begin{equation}\label{a3}
\hr(k)=\fr{\hk-\hk'}{\hk+\hk'},\qquad \htt(k)=\fr{2\hk}{\hk+\hk'}.
\end{equation}
 It is seen that the spinor \eref{a2} can be represented as
$|\Psi_0(x)\rangle=\hPs_0(x)|\xi\rangle$, where $\hPs_0(x)$ is
$2\times2$ matrix
\begin{equation}\label{a4}
\hPs_0(x)=\Theta(x<0)[\exp(i\hk x)+\exp(-i\hk
x)\hr(k)]+\Theta(x>0)\exp(i\hk' x)\htt(k).
\end{equation}
 This matrix satisfies the non-perturbed \Sc
\begin{equation}\label{a5}
[d^2/dx^2+k^2-B\sigma_z-u\Theta(x>0)]\hPs_0(x)=0,
\end{equation}
and is diagonal one, which means that its non-diagonal matrix
elements
\begin{equation}\label{a5a}
\hPs_0(x)_{du}=\langle\xi_d|\hPs_0(x)|\xi_u\rangle,\qquad
\hPs_0(x)_{ud}=\langle\xi_u|\hPs_0(x)|\xi_d\rangle
\end{equation}
are zero.

 The perturbation $b(x)\sigma_x$ changes \eref{a2}, and the change
$|\delta\Psi(x)\rangle=\delta\hPs(x)|\xi\rangle$, according to
perturbation theory is representable as
\begin{equation}\label{a6}
\delta\hPs(x)=\int\hG(x,x')b(x')\sigma_x\hPs_0(x')dx',
\end{equation}
where the matrix Green function, $\hG$, in the DWBA approach is a
causal solution of the inhomogeneous \Sc:
\begin{equation}\label{a7}
[d^2/dx^2+k^2-B\sigma_z-u\Theta(x>0)]\hG(x,x')=\hI\delta(x-x').
\end{equation}
 Here $\hI$ in the right hand side is the unit matrix.
 Solution of this equation is constructed with the help of two linearly
independent solutions $\hPs_{1,2}(x)$ of \eref{a5}:
\begin{equation}\label{a8}
\hG(x,x')=\hw^{-1}[\Theta(x>x')\hPs_1(x)\hPs_2(x')+
\Theta(x<x')\hPs_2(x)\hPs_1(x')],
\end{equation}
where $\hw$ is their Wronskian
\begin{equation}\label{a9}
\hw=[\hPs'_1(x)\hPs_2(x)-\hPs'_2(x)\hPs_1(x)],
\end{equation}
and prime means derivative over $x$: $\hPs'(x)=d\hPs(x)/dx$.

 For $\hPs_1(x)$ we can take solution \eref{a4}:
$\hPs_1(x)=\hPs_0(x)$, and for linear independent solution
$\hPs_2(x)$ we can take
\begin{equation}\label{a10}
\hPs_2(x)=\Theta(x<0)\exp(-i\hk x)\htt'(k)+\Theta(x>0)[\exp(-i\hk'
x)+\exp(i\hk x)\hr'(k)],
\end{equation}
where matching conditions satisfy for
\begin{equation}\label{a3a}
\hr'(k)=\fr{\hk'-\hk}{\hk+\hk'}=-\hr(k),\qquad
\htt'(k)=\fr{2\hk'}{\hk+\hk'}.
\end{equation}
The Wronskian of $\hPs_{1,2}(x)$ is equal to
\begin{equation}\label{a11}
\hw=2i\hk\htt'(k).
\end{equation}

 Substitution of all these matrices into \eref{a6} gives
$\delta\hPs(x)=\exp(-i\hk x)\hR$, where
\begin{equation}\label{a6a}
\hR= \int\limits_{-\infty}^0\fr{dx'}{2i\hk}[e^{i\shk x'}+e^{-i\shk
x'}\hr(k)]b(x')\sigma_x[e^{i\shk x'}+e^{-i\shk x'}\hr(k)].
\end{equation}
 The searched amplitude of spin flip reflection is
\begin{equation}\label{a12}
\hR_{du}=\langle\xi_d|\hR|\xi_u\rangle,
\end{equation}
 Substitution of \eref{a6a} gives
\begin{equation}\label{a12a}
\hR_{du}=\int\limits_{-\infty}^0\fr{dx'}{2ik_d}[e^{ik_d
x'}+e^{-ik_d x'}r_d(k)]b(x')[e^{ik_u x'}+e^{-ik_u x'}r_u(k)],
\end{equation}
where $k_{u,d}=\sqrt{k^2\mp B}$,
\begin{equation}\label{a14}
r_{u,d}(k)=\fr{k_{u,d}-k'_{u,d}}{k_{u,d}+k'_{u,d}},\qquad
k_{u,d}=\sqrt{k^2\mp B},\qquad k'_{u,d}=\sqrt{k^2-u\mp B}.
\end{equation}
 Substitution of $b(x)=b_0\exp(qx)$, where $q=1/\Lambda$, and
integration over $x'$ in \eref{a12a} gives
\begin{equation}\label{a15}
\hR_{du}=\fr{b_0}{2ik_d}\lt(\fr{1}{q+i(k_u+k_d)}+
\fr{r_ur_d}{q-i(k_u+k_d)}+\rt.$$
$$+\lt.\fr{r_d}{q+i(k_u-k_d)}+\fr{r_u}{q-i(k_u-k_d)}\rt)=$$
$$=\fr{b_0}{2ik_d}\lt(\fr{q(1+r_ur_d)-i(k_u+k_d)(1-r_ur_d)}{q^2+(k_u+k_d)^2}+
\rt.$$
$$+\lt.\fr{q(r_d+r_u)-i(k_u-k_d)(r_d-r_u)}{q^2+(k_u-k_d)^2}\rt).
\end{equation}
 In the case of total reflection and not large external field
($U_{magn}=|\mu|(\si\cdot\B)\ll E_{n}$)
we can approximate $k_u+k_d\approx2k$, $k_u-k_d\approx-B/k$ and
$r_u\approx r_d=\exp(-2i\phi)$, where $\phi=\arccos(k/\sqrt{u})$.
 In this approximation the Eq.~\eref{a15} is reduced to
\begin{equation}\label{a16}
\hR_{du}=e^{-2i\phi}\fr{b_0}{ik}\lt(\fr{q\cos(2\phi)+2k\sin(2\phi)}{q^2+4k^2}+
\fr{q}{q^2+B^2/k^2}\rt).
\end{equation}

 As we are interested in the interaction range satisfying to $k\Lambda\gg 1$
(typical UCN $k\sim 10^{6}\,cm^{-1},\, \Lambda>10^{-4}\,cm$) and
not too strong external magnetic fields ($<\sim 500$ G),
the first term in Eq. (23) can be neglected comparing to the second one:
\begin{equation}\label{a17}
\hR_{du}=e^{-2i\phi}\fr{b_0}{ik}\lt(\fr{q}{q^2+B^2/k^2}\rt),
\end{equation}
and the spin-flip reflectivity from the wall, represented in
dimensional units, becomes
\begin{equation}
w=|\hR_{du}|^{2}=\frac{4V_{0}^{2}<v_{\perp}>^{2}(1-e^{-d/\Lambda})^{2}}
{\Lambda^{2}\hbar^{2}\Biggl(\frac{<v_{\perp}>^{2}}{\Lambda^{2}}
+\omega_{0}^{2}\Biggr)^{2}},
\end{equation}
where $\omega_{0}=\gamma_{n}B$ is the neutron spin Larmor
frequency in the external field $B$,
$\gamma_{n}=1.83\times 10^{4}$\,s$^{-1}$/G - the gyromagnetic
ratio for the neutron, $<v_{\perp}>$ is the averaged over the UCN
spectrum normal to the surface neutron velocity component.

 At $\omega_{0}\gg <v_{\perp}>/\Lambda$ we have the expression coinciding
with the one obtained for the quasiclassical case and derived in the Ref.
\cite{dep} with $\omega_{0}^{-4}$ behavior of the depolarization probability.
 Typically $<v_{\perp}>\sim 300$ cm/s, and for $\Lambda=10^{-4}$ cm the
quasiclassical case is valid at $B>100$ G.

 At a weak guiding magnetic field $\sim 10^{-2}$ G the quasiclassical
approach is valid only for the interaction range
$\Lambda><v_{\perp}>/\omega_{0}\sim 1$ cm.

 Substituting (2) into (25) we obtain the expression for $g_{s}g_{p}$:
\begin{equation}
g_{s}g_{p}\simeq\beta^{1/2}\frac{2\,m_{n}} {\hbar
N<v_{\perp}>(1-e^{-d/\lambda})}
\Biggl(\frac{<v_{\perp}>^{2}}{\lambda^{2}}+\omega_{0}^{2}\Biggr),
\end{equation}
 where $\beta$ is the experimentally measured UCN depolarization probability
per one reflection from the wall of the storage cavity.

 There are two published experimental data on the ultra-cold neutron
depolarization: \cite{Serdep} and \cite{PSIdep}, in which special
experiments are described to measure this value.
 In both publications, for a variety of materials, the measured values of
the neutron depolarization probability per one neutron collision with
the walls of storage cavity were around $\beta\sim 10^{-5}$.
 The $\beta\sim 10^{-6}$ was measured in \cite{PSIdep} for the diamond like
carbon foils (DLC).
 The magnetic fields in the storage chambers were partly due to stray fields
from strong magnets used for the polarization of the incident
neutrons \cite{Serdep,PSIdep} and partly were formed by special
magnets \cite{PSIdep}. In \cite{Serdep} magnetic field was
estimated~\cite{Serinf} to be near 50 G , and in the
experiment~\cite{PSIdep} it was reported to be $\sim$55 G.
In both cases, this rather large external magnetic field suppressed
effect of the additional hypothetical spin-dependent interaction
on depolarization of the ultra-cold neutrons in traps, and
therefore decreased sensitivity of these measurements to
establishing constraints on the axion-like interaction.

A better constraints can be obtained from the measurements of the
ultra-cold neutron depolarization in traps at lower magnetic field
$B_{z}$ in the experiments on the search for the neutron electric
dipole moment (EDM) \cite{SerEDM} and \cite{ILLEDM}.
 There the ultra-cold neutrons preliminary polarized by transmission
 through magnetized ferromagnetic foil were stored in a cylindrical
 bottle permeated by magnetic and electric fields.
 The magnetic field was applied parallel to the
axis of the bottle, and its value in these experiments was very
low: $B_{z}=0.02$\,G in\cite{SerEDM}, and $B_{z}=0.01$\,G in
\cite{ILLEDM}.
 The change of the magnetic resonance frequency was
sought for at the reverse of the electric field direction.
 After filling the bottle with ultra-cold neutrons and closing the neutron
valve, the $\pi/2$ Ramsey pulse was applied, which turned neutron spins
perpendicular to the magnetic field.
 The neutrons were allowed to precess about magnetic field for $130$\,s
\cite{ILLEDM}, after which the second $\pi/2$ Ramsey pulse was
applied, and neutrons in the appropriate spin state passed back
through the polarizing foil to the neutron detector.

 Depolarization of neutrons at reflections from the walls
of the storage cavity in presence of a gradient of a
spin-dependent potential decreases contrast of the neutron
magnetic resonance curves.
 Probability of the neutron depolarization was not measured directly
 in these experiments, but from the reported very good magnetic resonance
 curves it can be concluded that the UCN depolarization probability at a single
collision with the walls was not higher than in \cite{Serdep,PSIdep}.
 According to \cite{Ivan} the neutron depolarization time in the EDM
experiment \cite{ILLEDM} can be estimated to be not less than
$\tau_{dep}\sim 800\,s$.
 From this figure we can estimate the
depolarization probability per single collision with the walls:
$\beta=l/(\tau_{dep}v)= 18/(800\times 500)\sim 4\times 10^{-5}$,
where $l$ is the mean free path between two consecutive
collisions, which can be found from dimensions of the neutron
bottle: in the experiment~\cite{ILLEDM} it was a cylinder of the
internal diameter about $44$\,cm and height $15$\,cm.
 We constrain the monopole-dipole interaction in two assumptions for
the depolarization probability in \cite{ILLEDM}: one --
$\beta=4\times 10^{-5}$, and another one -- corresponding to the
best value obtained in the experiments of the PSI group
\cite{PSIdep} -- $\beta=10^{-6}$.

 With above $\beta$ we can draw the limiting curves for the parameters
of the monopole-dipole coupling of the axion field, which is shown in Fig. 1
together with results from other publications.

 The ultra-cold neutron depolarization data may be used also to set
limits on the monopole-dipole coupling between neutrons and
electrons of the walls of the storage chambers.
 However, because density of the electrons in the medium is approximately
two times lower than the density of nucleons, the constraints are
respectively two times less strong.

\newpage

\begin{figure}
\begin{center}
\includegraphics[width=\textwidth]{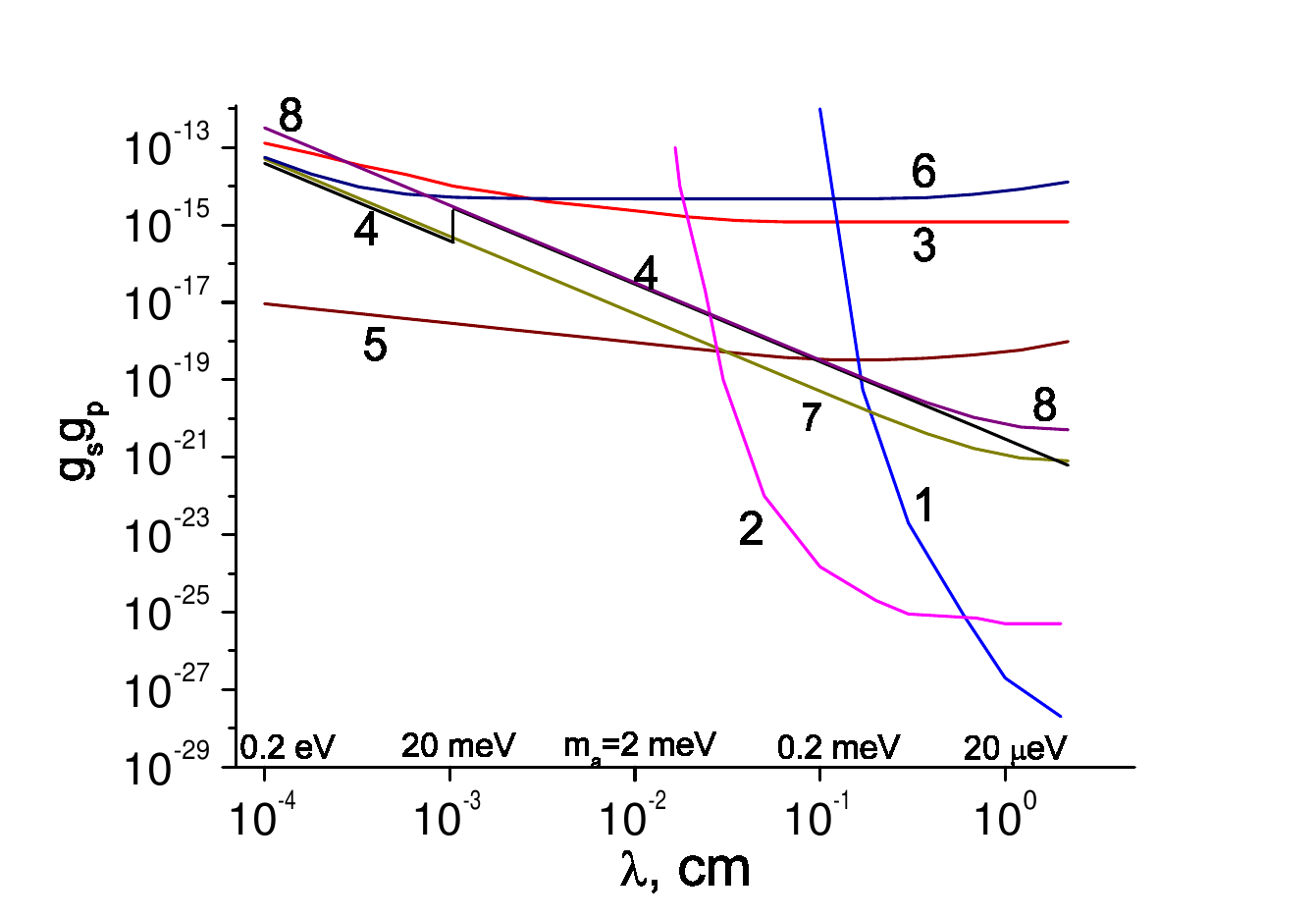}
\end{center}
\caption{Constraints on the axion monopole-dipole coupling
strength $g_{s}g_{p}$ and effective range:
1 and 2 -  constraints for the value of coupling constant of nucleon
and electron $g_{s}g_{p}^{e}$ from Refs. \cite{Ritt} and \cite{Hamm},
respectively;
3 - from the neutron gravity experiment of Ref. \cite{grav},
4 - from Ref. \cite{Ser},
5 - from spin relaxation of $^{3}He$, Ref. \cite {myHe},
6 - this work at the UCN depolarization probability $\beta=10^{-6}$ and
magnetic field $B_{z}=50$\,G \cite{PSIdep},
7 - the same, but $B_{z}=0.01\,G$ \cite{ILLEDM};
8 - the same, but $B_{z}=0.01\,G$, $\beta=4\times 10^{-5}$\cite{ILLEDM,Ivan}.
It was assumed in both cases of the ultra-cold neutron storage, that
$d=1\,cm$, $N\approx 2\times 10^{24}$\,cm$^{-3}$, $<v_{\perp}>=300$\,cm/s.}
\end{figure}

\end{document}